\documentstyle[prl,aps,multicol,epsf]{revtex}

\renewcommand{\d}{\displaystyle}
\newcommand{\ex}[1]{{\rm e}^{{\rm i}#1}}
\newcommand{\ve}{\varepsilon}
\newcommand{\vf}{v_{\rm F}}
\newcommand{\kf}{k_{\rm F}}
\begin{document}
\draft

\title{Tunneling exponents in realistic quantum wires using the mean 
field approximation}
\author{Wolfgang H\"ausler$^1$ and A.H. MacDonald$^2$}
\address{$^1$ Physikalisches Institut, Universit\"at Freiburg,
Hermann-Herder-Str. 3, D-79104 Freiburg, Germany\\
$^2$ Department of Physics, University of Texas at Austin, Austin TX, 78712,
U.S.A.}
\maketitle
\begin{abstract}
Article appeared in J.\ Phys.\ Soc.\ Jpn.\ {\bf 72}, 195 (2003).
\end{abstract}

\pacs{One-dimensional electrons, Quantum wires, Luttinger liquids.}
\begin{multicols}{2}
\narrowtext

Interacting carriers in one dimension (1D) are non-Fermi liquids
with power laws for many correlations functions such as the
tunneling density of states. Evidence for this behavior has
been found in carbon nanotubes \cite{bockrath} and, 
somewhat less convincingly, in semi conducting quantum wires
\cite{semicond} in the form of non-trivial temperature and transport
voltage dependences. Non-universal exponents are expressed through one
parameter $K_\rho$ within the Tomonaga-Luttinger (TL) liquid
theory.

In Fermi-liquids $K_\rho$ equals unity. This quantity is commonly
expected to decrease with increasing repulsion $V_0$ between the
carriers, according to
\begin{equation}\label{pop}
K_\rho=[1+2V_0/\pi v_{\rm F}]^{-1/2}\;.
\end{equation}
Eq.~(\ref{pop}) results also from the RPA-approximation for the
1D-plasmon velocity \cite{dassarma} which has shown to be exact
for spinless carriers and for strictly linear kinetic energy
dispersion \cite{larkin}. Eq.~(\ref{pop}) implies that
$K_{\rho}\to 0$ with decreasing carrier density
$n=2\kf/\pi=2m\vf/\pi$ for quadratic kinetic energy dispersion
$\sim k^2/2m$ when $\vf=\kf/m$ \cite{cntcomm}.

For given microscopic interaction potential $V(x)$, where the
Coulomb form in any realistic sample lay-out will be screened by
the nearest metals at a distance $R$, does $K_\rho$ depend on
the carrier density in a non-monotonous fashion, passing through
a minimum before reaching the asymptotic value which was
conjectured to be $K_\rho(n\to 0)\to 1/2$ \cite{whlkahm}.
Observing this minimum could give direct experimental access to
the range of the microscopic interaction. We shall argue that
the value of $K_\rho$ can be obtained quite accurately by the
Hartree-Fock approximation when augmenting self consistency
(SCHF).

As a realistic form modeling the microscopic interaction in a
quantum wire of width $d$ at particle separation $|x-x'|$ we use
\begin{equation}\label{interaction}
V(|x|)=\frac{e^2}{\epsilon}(\frac{1}{\sqrt{x^2+d^2}}-
\frac{1}{\sqrt{x^2+d^2+4R^2}})
\end{equation}
with $V(|x-x'|\gg R)\sim|x-x'|^{-3}$, describing dipolar
screening. We have solved the Hartree-Fock equations for a
system of length $L$, accounting for the quadratic kinetic
energy dispersion and for spin $s=\pm$

\begin{eqnarray}\label{schrod}
&&\!\!\!\!\!\!\!\!\!
0=\d\left[\frac{1}{2}(2j-\frac{k}{k_{\rm F}})^2
-\frac{\ve_{ks}}{\kf\vf}\right]u_{j,k,s}\\ \nonumber
&&\!\!\!\!\!\!\!\!\!
{}+\d\frac{L}{2k_{\rm F}\pi}\sum_{j'j''}u_{j'',k,s}
\int_{-k_{\rm F}}^{k_{\rm F}}{\rm d}k'
\Biggl\{\hat{V}\Bigl(2(j-j'')\Bigr)\\ \nonumber
&&\!\!\!\!\!\!\!\!\!\!\!\!\!\!\!\!\!
{}\times\d\sum_{s'}u_{-j+j'+j'',k',s'}^*\;u_{j',k',s'}\\ \nonumber
&&\!\!\!\!\!\!\!\!\!\!\!\!\!\!\!\!\!
{}-\d\hat{V}\Bigl(2(j-j')-\frac{k}{k_{\rm F}}+
\frac{k'}{k_{\rm F}}\Bigr)\:u_{-j+j'+j'',k',s}^*\;u_{j',k',s}\Biggr\}
\end{eqnarray}

\noindent self consistently in $k$-space for the coefficients
$u_{j,k,s}$ ($-\kf\le k\le\kf$) that expand the HF-orbitals
(index $j$) as Bloch waves
\begin{equation}\label{bloch}
\psi_{ks}(x)=\ex{kx}\sum_ju_{j,k,s}\:\ex{j2\kf x}\;.
\end{equation}

\begin{figure}
\epsfxsize=0.8\columnwidth
\centerline{\epsffile{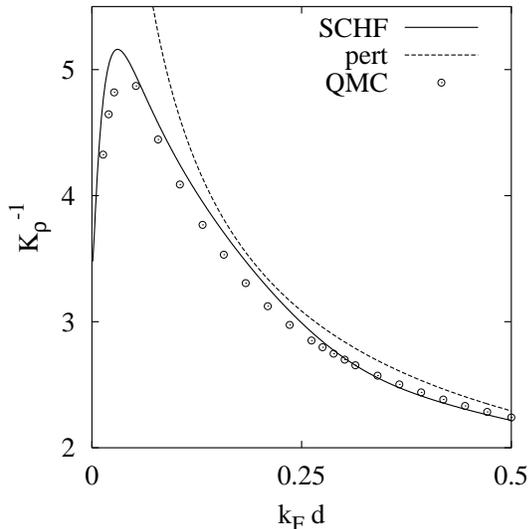}}
\caption[]{$K_{\rho}^{-1}$ versus carrier density. The units can
be translated into the $r_{\rm s}$-Fermi liquid parameter using
$\:r_{\rm s}=\pi/8\kf d\:$. The range of the microscopic
interaction (\ref{interaction}) is $R/d=14.1$. Solid: self
consistent HF approximation, dashed: Eq.~(\ref{pop}). QMC-data
are taken from Ref.~\cite{creffield}.}
\label{krho}
\end{figure}

Resulting total ground state energy densities $E_0^{\rm HF}/L$
are differentiated twice w.r.t.\ $n$ to obtain the HF-estimate
to the compressibility $\kappa=[\partial^2(E_0/L)/\partial
n^2]^{-1}$. Using the exact thermodynamic relationship
\begin{equation}\label{compress}
K_\rho=\sqrt{\pi\vf\kappa/2}
\end{equation}
from TL-theory \cite{haldane} yields $1/K_{\rho}^{\rm HF}$,
shown in Fig.~\ref{krho}. Also included in Fig.~\ref{krho} are
quantum Monte Carlo data taken from Ref.~\cite{creffield} that,
within symbol size, can be regarded as exact. It is seen that
$K_{\rho}^{\rm HF}$ does reproduce all of the available QMC-data
points amazingly well. In view of the pronounced correlations
of interacting one-dimensional Fermions, which prohibit for
example to express the ground state wave function analytically,
such a quite satisfying mean-field approach might seem
unexpected.
\newpage

The following general trends are seen in Fig.~\ref{krho}:\\
({\em i}) The high density region $\kf d\gtrsim 0.25$,
corresponding to $r_{\rm s}\lesssim 1.6$, may be regarded as the
perturbational or RPA regime. Here Eq.\ (\ref{pop}) may be
improved slightly by accounting for the exchange contribution
$\sim-\hat V(2\kf)$. Despite of the quite small values of
$K_{\rho}$ estimated \cite{egger} and observed \cite{bockrath}
in carbon they nanotubes belong typically to this regime, since
mean carrier separations exceed by far the interaction range
(which can reach the order of the tube length).\\
({\em ii}) Between $0.1\lesssim \kf d\lesssim 0.25$ the
perturbational expression still allows to guess $K_{\rho}$.
Here, particularly the SCHF but also the QMC-data indicate
slightly enhanced $K_{\rho}^{-1}$-values, relative to Eq.\
(\ref{pop}). By virtue of (\ref{compress}) this suggests a {\em
reduced} compressibility which can be interpreted as precursor
to a Wigner crystal phase transition (that cannot be completed
in 1D). There, $K_{\rho}$ has been estimated first in
\cite{glazman92c}.\\
({\em iii}) Finally, for the interaction range $R/d=14.1$ shown
in Fig.~\ref{krho}, a maximum is seen below $\kf d\lesssim 0.1$
($r_{\rm s}\gtrsim 4$) in both, the SCHF and the QMC results, in
qualitative difference to the monotonous increase of the
perturbative expression (\ref{pop}). This maximum occurs
roughly when $\kf R\approx\pi/8$ ($R$ has to be significantly
smaller than the mean carrier spacing due to quantum
fluctuations). In semi conducting quantum wires this regime
({\em iii}) should be feasible.

It has been conjectured that $1/K_{\rho}$ would approach 2 when
$\kf\to 0$ \cite{whlkahm}. This limit is not confirmed by the
SCHF which, when carried out carefully to account for the
pronounced $4\kf$-periodic oscillations of the SCHF-density
(resembling a Wigner crystal), yields $1/K_{\rho}^{\rm HF}\to
3.3$ with a weak dependence on $R/d$, ranging from
$1/K_{\rho}^{\rm HF}\to 3.0$ at $R/d=5.7$ to $1/K_{\rho}^{\rm
HF}\to 3.5$ at $R/d=35$. Unfortunately, the QMC-data cannot
discriminate between $1/K_{\rho}\to 2$ and $1/K_{\rho}\to 3.5$.
This discrepancy might indicate the failure of the mean field
approximation to estimate $K_{\rho}$ in the strongly correlated
regime of very small carrier densities.

We would like to acknowledge useful discussions with Charles
Creffield and with Hermann Grabert.

\begin{raggedright}

\end{raggedright}
\end{multicols}

\begin{thebibliography}{99}
\bibitem{bockrath} M. Bockrath, {\em et al.}, Nature {\bf 397} (1999) 598.
\bibitem{semicond}
 S. Tarucha, T. Honda, and T. Saku, Solid State Comm. {\bf 94} (1995) 413;
 A. Yacoby, {\em et al.}, Phys.\ Rev.\ Lett.\ {\bf 77} (1996) 4612;
 O.M. Auslaender, {\em et al.}, Phys.\ Rev.\ Lett.\ {\bf 84} (2000) 1764;
 M. Rother, {\em et al.}, Physica~E {\bf 6} (2000) 551.
\bibitem{dassarma} Q.P. Li, S. Das Sarma, and R. Joynt, Phys.\ Rev.\ B
 {\bf 45} (1992) 13713.
\bibitem{larkin} I.E. Dzyaloshinski\u{\i} and A.I. Larkin, Sov.\ Phys.\ JETP,
 {\bf 38} (1974) 202.
\bibitem{cntcomm} In systems with strict linear dispersion, such
 as carbon nanotubes, the value of the $K_\rho$-parameter cannot
 be regulated by varying the carrier density. This case is not
 investigated here.
\bibitem{whlkahm} W. H\"ausler, L. Kecke, and A.H. MacDonald,
 Phys.\ Rev.\ B {\bf 65} (2002) 085104.
\bibitem{haldane} F.D.M. Haldane, J.\ Phys.\ C {\bf 14} (1981) 2585.
\bibitem{creffield} C.E. Creffield, W. H\"ausler, and A.H. MacDonald,
 Europhys.\ Lett.\ {\bf 53} (2001) 221.
\bibitem{egger} R. Egger and A.O. Gogolin, Eur.\ Phys.\ J.\ B
 {\bf 3} (1998) 281.
\bibitem{glazman92c} L. I. Glazman, I. M. Ruzin, and B. I. Shklovski\u{\i},
 Phys.\ Rev.\ B {\bf 45} (1992) 8454.
\end{thebibliography}
\end{document}